\documentclass[apj]{emulateapj}
\usepackage{graphics,graphicx,times,natbib,longtable,placeins}

\def\mhalo{${\rm M_{halo}}$}
\def\mstar{${\rm M_{*}}$}
\def\msun{${\rm\,M_\odot}$}

\def\tq{$T_{Q}$}
\def\dqprojrvir{$d_{proj}^Q / R_{Vir}$}

\def\spose#1{\hbox to 0pt{#1\hss}}
\def\lta{\mathrel{\spose{\lower 3pt\hbox{$\mathchar"218$}}
     \raise 2.0pt\hbox{$\mathchar"13C$}}}

\shorttitle{Environments of Dwarf Galaxies}
\shortauthors{Guo et al.}

\begin{document}

\title{CANDELS Sheds Light on the Environmental Quenching of Low-mass Galaxies}

%\author{Yicheng Guo$^{1}$, et al.}
%\affil{$^1$ UCO/Lick Observatory, Department of Astronomy and Astrophysics, University of California, Santa Cruz, CA, USA; {\it ycguo@ucolick.org}}
\author{Yicheng Guo\altaffilmark{1}}
\author{Eric F. Bell\altaffilmark{2}}
\author{Yu Lu\altaffilmark{3}}
\author{David C. Koo\altaffilmark{1}}
%%%%%%%%%%%%%%
\author{S. M. Faber\altaffilmark{1}}
\author{Anton M. Koekemoer\altaffilmark{4}}
\author{Peter Kurczynski\altaffilmark{5}}
\author{Seong-Kook Lee\altaffilmark{6}}
\author{Casey Papovich\altaffilmark{7,8}}
%%%%%%%%%%%%%%
%\author{Nimish P. Hathi\altaffilmark{7}}
\author{Zhu Chen\altaffilmark{9}}
\author{Avishai Dekel\altaffilmark{10}}
\author{Henry C. Ferguson\altaffilmark{4}}
\author{Adriano Fontana\altaffilmark{11}}
\author{Mauro Giavalisco\altaffilmark{12}}
\author{Dale D. Kocevski\altaffilmark{13}}
\author{Hooshang Nayyeri\altaffilmark{14}}
\author{Pablo G. P\'erez-Gonz\'alez\altaffilmark{15}}
\author{Janine Pforr\altaffilmark{16,17}}
\author{Aldo Rodr\'iguez-Puebla\altaffilmark{18}}
\author{Paola Santini\altaffilmark{11}}
\altaffiltext{1}{UCO/Lick Observatory, Department of Astronomy and Astrophysics, University of California, Santa Cruz, CA, USA; {\it ycguo@ucolick.org}}
\altaffiltext{2}{Department of Astronomy, University of Michigan, Ann Arbor, MI, USA}
\altaffiltext{3}{Observatories, Carnegie Institution for Science, Pasadena, CA, USA}
\altaffiltext{4}{Space Telescope Science Institute, Baltimore, MD, USA}
\altaffiltext{5}{Department of Physics and Astronomy, Rutgers University, Piscataway, NJ, USA}
\altaffiltext{6}{Center for the Exploration of the Origin of the Universe, Department of Physics and Astronomy, Seoul National University, Seoul, Korea}
\altaffiltext{7}{Department of Physics and Astronomy, Texas A\&M University, College Station, TX, USA}
\altaffiltext{8}{George P. and Cynthia Woods Mitchell Institute for Fundamental Physics and Astronomy, Texas A\&M University, College Station, TX, USA}
%\altaffiltext{8}{Aix Marseille Universit{\'e}, CNRS, LAM (Laboratoire d'Astrophysique de Marseille) UMR 7326, Marseille, France}
\altaffiltext{9}{Shanghai Key Lab for Astrophysics, Shanghai Normal University, 100 Guilin Road, 200234, Shanghai, China}
\altaffiltext{10}{Center for Astrophysics and Planetary Science, Racah Institute of Physics, The Hebrew University, Jerusalem, Israel}
\altaffiltext{11}{INAF-Osservatorio Astronomico di Roma, Via Frascati 33, I-00078 Monte Porzio Catone, Rome, Italy}
\altaffiltext{12}{Department of Astronomy, University of Massachusetts, Amherst, MA, USA}
\altaffiltext{13}{Colby College, Waterville, ME, USA}
\altaffiltext{14}{Department of Physics and Astronomy, University of California, Irvine, CA, USA}
\altaffiltext{15}{Departamento de Astrof\'{\i}sica, Facultad de CC.  F\'{\i}sicas, Universidad Complutense de Madrid, E-28040 Madrid, Spain}
\altaffiltext{16}{Aix Marseille Universit{\'e}, CNRS, LAM (Laboratoire d'Astrophysique de Marseille) UMR 7326, 13388, Marseille, France}
\altaffiltext{17}{ESA/ESTEC, Noordwijk, The Netherlands}
\altaffiltext{18}{Instituto de Astronom\'ia, Universidad Nacional Aut\'onoma de M\'exico, A. P. 70-264, 04510 M\'exico, D.F., M\'exico}
%\altaffiltext{19}{INAF - Osservatorio Astronomico di Roma, via Frascati 33, 00078, Monteporzio, Italy}

%\ \\

%\date{{\sc Version 1c: } \today }

\begin{abstract} 

We investigate the environmental quenching of galaxies, especially those with
stellar masses (\mstar)$<10^{9.5}$\msun, beyond the local universe. Essentially
all local low-mass quenched galaxies (QGs) are believed to live close to
massive central galaxies, which is a demonstration of environmental quenching.
We use CANDELS data to test {\it whether or not} such a dwarf QG--massive
central galaxy connection exists beyond the local universe. For this purpose, we
only need a statistically representative, rather than a complete, sample of
low-mass galaxies, which enables our study to $z\gtrsim1.5$. For each low-mass
galaxy, we measure the projected distance ($d_{proj}$) to its nearest massive
neighbor (\mstar$>10^{10.5}$\msun) within a redshift range. At a given $z$ and
\mstar, the environmental quenching effect is considered to be observed if the
$d_{proj}$ distribution of QGs ($d_{proj}^Q$) is significantly skewed toward
lower values than that of star-forming galaxies ($d_{proj}^{SF}$). For galaxies
with $10^{8}$\msun$<$\mstar$<10^{10}$\msun, such a difference between
$d_{proj}^Q$ and $d_{proj}^{SF}$ is detected up to $z\sim1$. Also, about 10\%
of the quenched galaxies in our sample are located between two and four virial
radii ($R_{Vir}$) of the massive halos. 
%implying that quenching starts from large halo radii. 
The median projected distance from low-mass QGs to their massive neighbors,
$d_{proj}^Q / R_{Vir}$, decreases with satellite \mstar\ at \mstar$\lesssim
10^{9.5}$\msun, but increases with satellite \mstar\ at \mstar$\gtrsim
10^{9.5}$\msun. This trend
%but not on redshift. $d_{proj}^Q / R_{Vir}$ decreases with satellite \mstar,
%Our results suggest Our results 
suggests a smooth, if any, transition of the quenching timescale around
\mstar$\sim 10^{9.5}$\msun\ at $0.5<z<1.0$.

%and (2) no change of quenching mechanism between $z \sim 1$ and $z=0$.} at
%$z\lesssim1$, but not found at $z>1.2$. The highest $z$ to which such a
%difference is detected slightly depends on \mstar, implying a gradually
%increase of the environmental quenching timescale (\tq) with \mstar\ of
%low-mass galaxies. Our method provides a novel, uniform way to constrain \tq\
%over two orders of magnitude in \mstar. Our results, together with other \tq\
%measurements in the literature, imply a transition of \tq\ (and therefore
%quenching mechanisms) at \mstar$\sim 10^{10}$\msun.

\end{abstract}

\section{Introduction}
\label{intro}

Environmental effects are believed to be the primary process of ceasing star
formation in low-mass galaxies with stellar masses (\mstar) lower than
$10^{9.5}$\msun\ (or dwarf galaxies). Field low-mass galaxies may temporarily
quench their star formation through supernova feedback, but new gas accretion
and recycling would induce new starbursts with periods of tens of megayears
\citep[e.g.,][]{ycguo16bursty,sparre17}. \citet[][hereafter G12]{geha12} found
that the quenched fraction of galaxies with \mstar$<10^{9}$\msun\ drops rapidly
as a function of distance to massive host galaxies and that essentially all
local field galaxies in this mass regime are forming stars.
%\citet{davies16} also found that basically all quenched \mstar$<10^{9}$\msun\
%galaxies are found in pairs or groups with a massive neighbor.

The environmental quenching of low-mass galaxies beyond the local universe,
however, is rarely investigated because of these galaxies' faint luminosity.
Most studies \citep[e.g., G12;][etc.]{quadri12, tal13, tal14, balogh16,
nancy16, fossati17} start from central galaxies and measure the quenched
fraction of their satellites. This method requires a complete sample of
satellites, which limits these studies to the local universe and/or to
intermediate-mass (\mstar$\gtrsim 10^{9.5}$\msun) satellites. 

In this letter, we use CANDELS data \citep{candelsoverview,candelshst} to
detect the effects of environmental quenching beyond $z\sim1$. Our
approach is different from but complementary to other studies. We start from
the ``victims'' --- quenched dwarf galaxies --- and search for their massive
neighbors, which are tracers of massive dark matter halos.

The concept of our approach is simple --- if environmental effects are solely
responsible for quenching all dwarf galaxies, all low-mass quenched galaxies
(QGs) should live close to a massive central galaxy in a massive halo. In
contrast, star-forming galaxies (SFGs) can live far away from massive dark
matter halos.  Therefore, on average, QGs should have systematically shorter
distances to their massive neighbors than SFGs should.  This systematic
difference between the two populations is evidence of the dwarf QG--massive
central galaxy connection and therefore a demonstration of environmental
quenching. Because our goal is to investigate {\it whether or not} such a dwarf
QG--massive central connection has been established, we only need to detect a
statistically meaningful signal, rather than to find all signals, to rule out
the null hypothesis of no environmental effects. This advantage allows us to
use an incomplete dwarf sample to study this topic beyond the local universe.

%We identify quenched and star-forming galaxies from CANDELS data through their
%rest-frame optical--NIR colors. For each galaxy, we measure the projected
%distance ($d_{proj}$) to its nearest massive neighbor
%(\mstar$>10^{10.5}$\msun) within a projected volume. In a given $z$ and
%\mstar\ bin, we compare the $d_{proj}$ distribution of quenched galaxies to
%that of star-forming galaxies. 

We adopt a flat ${\rm \Lambda CDM}$ cosmology with $\Omega_m=0.3$,
$\Omega_{\Lambda}=0.7$, and the Hubble constant $h\equiv H_0/100\ {\rm
km~s^{-1}~Mpc^{-1}}=0.70$. We use the AB magnitude scale \citep{oke74} and a
\citet{chabrier03} IMF.

\begin{figure*}[htbp]
%\center{\hspace*{-0.0cm}\includegraphics[scale=0.30,
%angle=0]{./proxpdf_0.50_0.75_8.0_8.5.ps}}
\hspace*{-0.55cm}\includegraphics[scale=0.22, angle=0]{./fig1a.ps}
\hspace*{-0.55cm}\includegraphics[scale=0.22, angle=0]{./fig1b.ps}
\hspace*{-0.55cm}\includegraphics[scale=0.22, angle=0]{./fig1c.ps}

\caption[]{Examples of sample selection and environment measurement. Each
column shows a given ($z$, \mstar) bin as the title shows. In each column,
Panel (a) shows the selected QGs (red) and SFGs (blue) in the UVJ diagram.
Black dots show all galaxies (with $H_{F160W}<26$ and CLASS\_STAR$<$0.8) in
this bin. Panel (b) shows the PDFs of $d_{proj}$ of the QGs (red) and SFGs
(blue) galaxies. The dark, medium, and light gray regions show the 1$\sigma$,
2$\sigma$, and 3$\sigma$ levels of 3000 times of bootstrapping of SFGs to match
the number of the QGs. The red and blue circles show the medians of $d_{proj}$
of the QGs and SFGs. The gray bar shows the 1$\sigma$ (dark), 2$\sigma$
(medium), and 3$\sigma$ (light) levels of the medians of the bootstrapping. To
show the difference clearly, all median values are normalized so that the
median of the SFGs (blue circle) is equal to 1.5 Mpc. The number below the gray
bar shows the confidence level to which the null hypothesis that the QGs (red)
and SFGs (blue) have the same $d_{proj}$ medians is ruled out.  Panel (c) shows
the CDFs of $d_{proj}$ of the QGs (red) and SFGs (blue) normalized by $R_{Vir}$
of the halos of their massive neighbors. The dark, medium, and light gray
regions show the 1$\sigma$, 2$\sigma$, and 3$\sigma$ levels of the
bootstrapping. All columns in this figure are at $0.5<z<0.75$.

\label{fig:example1}}
%\vspace{-0.2cm}
\end{figure*}

\begin{figure*}[htbp]
%\center{\hspace*{-0.0cm}\includegraphics[scale=0.30, angle=0]{./proxpdf_0.75_1.00_8.5_9.0.ps}}
%\vspace*{-1.55cm}
\hspace*{-0.55cm}\includegraphics[scale=0.22, angle=0]{./fig2a.ps}
\hspace*{-0.55cm}\includegraphics[scale=0.22, angle=0]{./fig2b.ps}
\hspace*{-0.55cm}\includegraphics[scale=0.22, angle=0]{./fig2c.ps}

\caption[]{Same as Figure \ref{fig:example1}, but showing three \mstar\ bins at
$0.75<z<1.00$. 

\label{fig:example2}}
%\vspace{-0.2cm}
\end{figure*}

\begin{figure*}[t!]
%\center{\hspace*{-0.0cm}\includegraphics[scale=0.30, angle=0]{./proxpdf_1.00_1.25_10.0_10.5.ps}}
\hspace*{-0.55cm}\includegraphics[scale=0.22, angle=0]{./fig3a.ps}
\hspace*{-0.55cm}\includegraphics[scale=0.22, angle=0]{./fig3b.ps}
\hspace*{-0.55cm}\includegraphics[scale=0.22, angle=0]{./fig3c.ps}

\caption[]{Same as Figure \ref{fig:example1}, but showing three \mstar\ bins at
$1.00<z<1.25$. 

\label{fig:example3}}
%\vspace{-0.2cm}
\end{figure*}

\section{Data}
\label{data}

We use the photometric redshift (photo-z), \mstar, and rest-frame color
catalogs of four CANDELS fields: GOODS-S \citep{ycguo13goodss}, UDS
\citep{galametz13uds}, GOODS-N (G. Barro et al., in preparation) and COSMOS
\citep{nayyeri17cosmos}.

The photo-z measurement is described in \citet{dahlen13}. 
%We use CANDELS/GOODS-S to test the photo-z accuracy. 
For GOODS-S galaxies at $0.5<z<2.0$ and
$H<26$ AB, the 1$\sigma$ scatter of $|\Delta z|/(1+z)$ is 0.026 and the outlier
fraction (defined as $|\Delta z|/(1+z)>0.1$) is 8.3\%.  We also divide the test
sample into low-mass (\mstar$<10^{9}$\msun) and massive (\mstar$>10^{9}$\msun)
sub-samples. The 1$\sigma$ scatter and outlier fraction of the low-mass (and
massive) sub-sample are 0.033 (0.024) and 13.7\% (7.2\%).

The \mstar\ measurement is described in \citet{santini15}, where each galaxy
is fit by 12 SED-fitting codes with different combinations of synthetic
stellar population models, star formation histories, fitting methods, etc. For
each galaxy, we use the median of the 12 best-fit \mstar\ as its \mstar.  The
typical uncertainty of \mstar\ measurement is $\sim$0.15 dex. Rest-frame colors
are measured by using EAZY \citep{brammer08}. 

\begin{figure}[htbp]
\center{\vspace*{-0.20cm}\hspace*{-0.0cm}\includegraphics[scale=0.17, angle=0]{./fig4.ps}}

\caption[]{{\it Panel (a)}: Statistics of the quenching--environment
connection.  In each ($z$, \mstar) bin, the deviation between the medians of
$d_{proj}^{Q}$ and $d_{proj}^{SF,sub}$ is the upper number, while the numbers
of the QGs and SFGs are the two lower numbers. We choose the deviation
$\geq$3$\sigma$ as the threshold of the quenching--environment connection being
observed. All such bins are red, while others with the deviation $<$3$\sigma$
are cyan. Gray bins cannot be accessed by our current dataset. {\it Panel (b)}: 
%Values of the CDFs (see Panel (c) of Figure 1--3) at $2 R_{Vir}$ of the halos
%of massive neighbors. For each sample (indicated by different lables), the
%value showes the 
Fraction of a population of galaxies within $2 R_{Vir}$ of massive halos.
Different symbols show different populations. {\it Panel (c)}: Median $d_{proj}
/ R_{Vir}$ of different samples. {\it Panel (d)}: Inferred quenching timescales
in two redshift bins. The colors and symbols in Panel (b), (c), and (d) are the
same.

%inferred quenching timescales. See Section \ref{results} for the calculation
%method. Error bars in both directions are calculated from the 16th and 84th
%percentiles of the low-mass galaxies. We also enlarge and shrink our redshift
%bin size to test the robustness of statistics, as different colors show. {\it
%Lower}: Comparison between our fiducial measurements (red symbols) with others
%in the literature.  We shift the data of middle panel by -0.28 dex in \mstar\
%to account for the mass loss since being quenched to $z=0$. Note that all our
%measurements should be treated as upper limits.}

\label{fig:spatial}}
%\vspace{-0.2cm}
\end{figure}

\section{Method}
\label{method}

%Figure \ref{fig:example1}, \ref{fig:example2}, and \ref{fig:example3} show
%three examples to summarize our sample selection, measurement, and analysis.

\subsection{Sample Selection}
\label{sub:sample}

Our sample consists of sources with F160W $H<26$ AB, PHOTFLAG=0 (no suspicious
photometry), and SExtractor CLASS\_STAR$<$0.8. The magnitude limit of $H=26$ AB
is approximately the 50\% completeness limit of CANDELS wide regions
\citep{ycguo13goodss} and it is corresponding to a galaxy of \mstar$\sim
10^{8}$\msun\ at $z \sim 0.5$ with a single stellar population that is 5 Gyr
old. We divided the whole sample into different $z$ and \mstar\ bins: $z =
0.5-2$ with $\Delta z=0.25$ and ${\rm log(M_{*})} = 8.0-10.5$ with ${\rm \Delta
log(M_{*})} = 0.5$.
%step size of 0.5 dex.  The redshift bins span between 0.5 and 2 with a bin
%size of 0.25, and the ${\rm log(M_{*})}$ bins span between 8.0 and 10.5 with a
%bin size of 0.5 dex.

In each ($z$, \mstar) bin, we use the UVJ diagram \citep{williams09,muzzin13}
to select QGs and SFGs. To avoid the contamination of
misidentified stars and sources with suspicious colors, we add one criterion
to refine the quenching region (the diagonal light brown line within the
original UVJ quenched region in Panel (a)s of Figures
\ref{fig:example1}--\ref{fig:example3}). This extra criterion may exclude some
very compact QGs \citep{barro13a}, but since our goal is to
obtain a clean and statistically meaningful sample rather than a complete one,
such exclusion is necessary and does not affect our results.

For SFGs, instead of using all galaxies in the UVJ star-forming region, we
measure the median and $\pm 1.5 \sigma$ level of the star-forming locus
(calculated in the directions parallel and perpendicular to the reddening
vector) and use them as the selection boundary. The selected SFGs are plotted
as blue points in Panel (a)s of Figures \ref{fig:example1}--\ref{fig:example3}.
Again, although many galaxies in the original UVJ star-forming locus are
excluded, we aim at constructing a clean rather than complete sample.

%\subsection{Finding the Neighbors}
\subsection{Detecting Environmental Quenching Effects}
\label{sub:neighbor}

For each low-mass galaxy, we search for its nearest massive neighbor in sky
(projected distance). The massive sample is selected to have CLASS\_STAR$<$0.8,
PHOTFLAG=0, \mstar$^{massive}>10^{10.5}$\msun, and
\mstar$^{massive}>$\mstar$^{low-mass}$+0.5 dex\footnote{The last requirement
only affects galaxies with
$10^{10}$\msun$<$\mstar$^{low-mass}$$<$$10^{10.5}$\msun. The \mstar$^{massive}$
threshold of our massive sample corresponds to dark matter halos of
\mhalo$\gtrsim10^{12}$\msun. Since the \mstar--\mhalo\ relation evolves little
with redshift in this mass regime \citep{behroozi13}, our choice of a fixed
\mstar$^{massive}$ threshold at different redshifts allows us to investigate
the environmental effects of similar \mhalo\ at different cosmic times.}. The
redshift range of the massive sample is limited to $|z_{massive} -
z_{low-mass}|/(1+z_{low-mass})<0.10$, which is about 3$\sigma$ of our photo-z
accuracy. We calculate the projected distances between the low-mass galaxy and
the selected massive galaxies. The massive galaxy with the smallest
projected distance is chosen as the central galaxy of the low-mass galaxy. We
use $d_{proj}$ to denote this smallest projected distance.

Because of projection effects, a massive neighbor found through this method may
not be the real massive galaxy whose dark matter halo was responsible for
quenching the low-mass galaxy. But if environmental effects are the primary way
of quenching a population of low-mass galaxies, 
%quenched galaxies should only be found close to massive dark matter halos,
%while star-forming galaxies can be found far away from massive halos.
%Therefore, 
statistically, QGs should be located closer to massive companions
than SFGs should. As a result, the $d_{proj}$ distribution of
a quenched population ($d_{proj}^Q$) should be skewed toward lower values than
that of SFGs ($d_{proj}^{SF}$).

Many studies \citep[e.g.,][]{scoville13,davies16} used the local overdensity
field constructed by Voronoi tessellation or the nearest neighbor method to
measure environments. Since the local overdensity of a satellite galaxy is
correlated with $d_{proj}$, our method is similar to those using a density
field. 
%Although the density field approach is expected to be more accurate than our
%method, 
While our simple method provides necessary information to test the {\it whether
or not} question of our particular interest, future work with the density field
approach could provide more accurate and detailed measurements of environmental
quenching.
%, especially when using photo-z probability distribution function
%\citep[e.g.,][]{davies16}, We therefore choose our methond and defer other
%measurements in future work.}

%We use two statistics to test whether $d_{proj}^Q$ is systematically and
%significantly smaller than $d_{proj}^{SF}$ in each ($z$, \mstar) bin. The
%first is the cumulative distribution functions (CDFs) of $d_{proj}$ of both
%populations. The K-S test needs to show a probability of $<$0.05 to exclude
%the null hypothesis that both populations have the same $d_{proj}$
%distributions.  The second is the medians of $d_{proj}$ of both populations.
%Because in most We use the median $d_{proj}$ of both populations to 
We test whether $d_{proj}^Q$ is systematically and significantly smaller than
$d_{proj}^{SF}$ in each ($z$, \mstar) bin. Because in most ($z$, \mstar) bins,
the number of QGs is much smaller than that of SFGs, the small number
statistics needs to be taken into account. In each of these bins, we randomly
draw a sub-sample of the SFGs to match the number of the QGs and calculate the
median, probability distribution function (PDF), and cumulative distribution
functions (CDF) of $d_{proj}$ of the sub-sample ($d_{proj}^{SF,sub}$). We
repeat this bootstrapping sampling 3000 times, obtaining 3000 distributions of
$d_{proj}^{SF,sub}$. To exclude the null hypothesis, we ask the median
$d_{proj}^{Q}$ to be 3$\sigma$ smaller than the median of $d_{proj}^{SF,sub}$. 
%We find the second test (median) is more strict than the first one (CDF).
%Therefore, we use the second test as our main criterion.

Panels (b) in Figures \ref{fig:example1}--\ref{fig:example3} show some examples
of our results. The ($z$, \mstar) bins labeled with $\geq$3$\sigma$ values
(i.e., median $d_{proj}^{Q}$ is 3$\sigma$ smaller than median
$d_{proj}^{SF,sub}$) are considered to have an established
quenching--environment connection.
%meaning statistically one difference between quenched and star-forming
%galaxies is their distance to massive companions. 
In contrast, bins with $<$3$\sigma$ values cannot rule out the null hypothesis
of the two populations having the same $d_{proj}$ distributions with 3$\sigma$
confidence. 

\begin{figure}[htbp]
\center{\vspace*{-0.20cm}\hspace*{-0.0cm}\includegraphics[scale=0.22, angle=0]{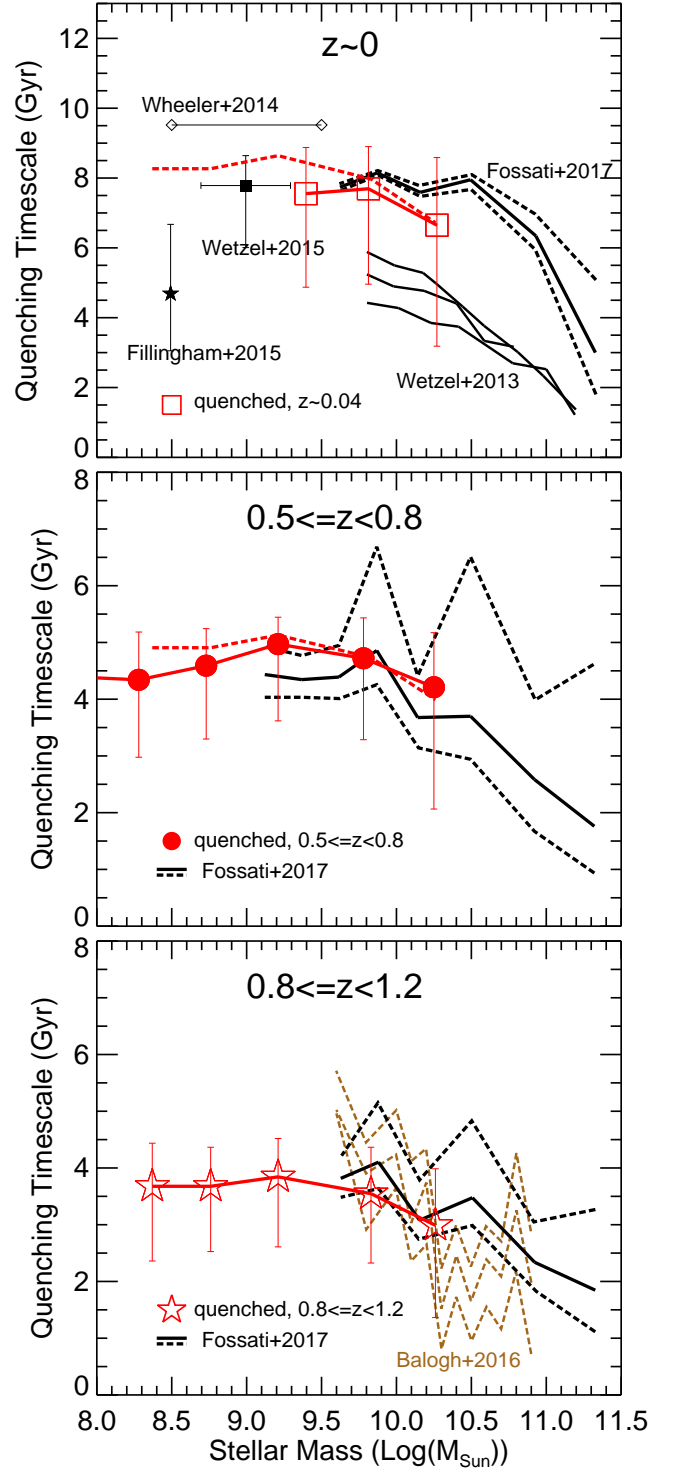}}

\caption[]{Quenching timescale at different redshifts. The red symbols are
calculated by this work, while the black and brown lines and symbols are taken
from the literature. The red dashed lines in the first two panels are the solid
red line in the third panel (i.e., \tq\ at $0.8\leq z<1.2$) scaled up by
$(1+z)^{1.5}$.

%{\it Panel (a)}: Statistics of the quenching--environment connection in all
%($z$, \mstar) bins. In each bin, the deviation between the medians of
%$d_{proj}^{Q}$ and $d_{proj}^{SF,sub}$ is shown as the upper number, while the
%numbers of the quenched and star-forming galaxies are shown as the two lower
%numbers. We choose the deviation $\geq$3$\sigma$ as the threshold of the
%quenching--environment connection being observed. All such bins are color
%coded by red, while others with the deviation $<$3$\sigma$ are cyan. Gray bins
%cannot be accessed by our current dataset.  {\it Middle}: inferred quenching
%timescales. See Section \ref{results} for the calculation method. Error bars
%in both directions are calculated from the 16th and 84th percentiles of the
%low-mass galaxies. We also enlarge and shrink our redshift bin size to test
%the robustness of statistics, as different colors show. {\it Lower}:
%Comparison between our fiducial measurements (red symbols) with others in the
%literature.  We shift the data of middle panel by -0.28 dex in \mstar\ to
%account for the mass loss since being quenched to $z=0$. Note that all our
%measurements should be treated as upper limits.

\label{fig:timescale}}
%\vspace{-0.2cm}
\end{figure}

\section{Results}
\label{results}

Panel (a) of Figure \ref{fig:spatial} shows the deviation from $d_{proj}^{Q}$
to $d_{proj}^{SF,sub}$ in each ($z$, \mstar) bin. Our criterion of
environmental quenching being observed is that the median of $d_{proj}^{Q}$ is
3$\sigma$ smaller than that of $d_{proj}^{SF,sub}$. For galaxies with
$10^8$\msun$<$\mstar$<10^{10}$\msun, such a quenching--environment connection
is observed up to $z \sim 1$, as shown by the larger-than-3$\sigma$ deviations.
This result is consistent with the quick emergence of low-mass QGs from the
measurement of stellar mass functions at $z\sim1$
\citep[e.g.,][]{ilbert13,huangjs13}. For galaxies with
$10^{10.0}$\msun$<$\mstar$<10^{10.5}$\msun, the connection was established at a
lower redshift.
%later time than lower-mass galaxies.

For those ($z$, \mstar) bins with $<$3$\sigma$ deviation, we cannot rule out
the null hypothesis of QGs and SFGs having the same $d_{proj}$ distributions
with more than 3$\sigma$ confidence. This may imply that the
quenching--environment connection has not been established in these bins. This
interpretation is at least consistent with some other studies for massive
galaxies (\mstar$>$$10^{10}$\msun), which claimed that massive quiescent
galaxies at $z>1$ are not necessarily located in high-density environments
\citep[e.g.,][]{darvish15,darvish16,lin16}. For lower-mass galaxies at
$z\gtrsim1.5$, however, due to projection effects or small number statistics,
our method may have failed to detect existing quenching--environment
connections.

%We also test the robustness of the statistics in three ways. First, we
%increase (and decrease) the width of redshift bins (within which we measure
%the medians of $d_{proj}^Q$ and $d_{proj}^{SF,sub}$) from $\Delta z = 0.25$ to
%0.30 (and to 0.20). Second, we increase the line-of-sight projection distance
%(within which we look for massive neighbors for each low-mass galaxy) from
%$|z_{massive} - z_{low-mass}|/(1+z_{low-mass}) < 0.10$ to $<0.20$. Third, to
%test the cosmic variance, we intently drop one CANDELS field in turn and
%re-calculate all the statistics. Our results are not significantly changed in
%all the tests.

\section{Discussion}
\label{discussion}

\subsection{Spatial Distribution of Quenched Galaxies}
\label{discussion:spatial}

%The spatial distribution of quenched galaxies around massive neighbors
%contains important information of quenching mechanisms. 
G12 found that 87\% (and 97\%) of dwarf QGs in their SDSS sample are within 2
$R_{Vir}$ (and 4 $R_{Vir}$) of a massive host galaxy. 
%while 97\% of objects are within 4 $R_{Vir}$. 
We find similar results in our sample at $0.5<z<1.0$ (Panel (b) of Figure
\ref{fig:spatial}). About 90\% of the QGs below $10^{10}$\msun\ in our sample
are within 2 $R_{Vir}$. The fraction drops quickly to about 70\% for galaxies
above $10^{10}$\msun. 
%In contrast, the fraction of SFGs within 2 $R_{Vir}$ is about 80\% below
%$10^{10}$\msun\ and drops to 70\% at $> 10^{10}$\msun. 
To calculate $R_{Vir}$, we first use the \mstar--\mhalo\ relation of
\citet{behroozi13} to obtain \mhalo\ ($M_{\rm vir}$) of the massive neighbors.
Then, we derive $R_{Vir}$ through $M_{\rm vir}(z) = {4\pi \over 3} \Delta_{\rm
c}(z) \rho_{\rm crit}(z) R_{\rm vir}(z)^3$, where $\rho_{\rm crit}(z)$ is the
critical density of the universe at $z$, and $\Delta_{\rm c}(z)$ is calculated
by following \citet{bryan98}. 

SFGs in our sample are also almost within 4 $R_{Vir}$. This could be a
projection effect. Since we search for massive neighbors within a long
line-of-sight distance ($|\Delta z|/(1+z)<0.10$), an SFG has a high chance of
being located within the projected 4 $R_{Vir}$ of a massive galaxy, even though
the massive galaxy is not its real central galaxy. In contrast, with a more
accurate redshift measurement, G12 found that only $\sim$50\% of the $z\sim0$
SFGs are within 4 $R_{Vir}$, suggesting that a large fraction of SFGs are
intrinsically outside 4 $R_{Vir}$. QGs in our sample may suffer from the same
projection effect.  However, as discussed in Section \ref{sub:neighbor}, this
effect would not affect our {\it statistical} results.

A non-negligible fraction (10\%) of low-mass QGs are located between 2 and 4
$R_{Vir}$. They are likely central galaxies quenched by mechanisms not related
to environment, e.g., AGN and stellar feedback. They, however, may also be
evidence for quenching processes acting at large distances of massive halos
\citep[e.g.,][]{cen14}. Y. Lu et al. (2017, in preparation) found that, to
match the observed \mstar\ and stellar-phase metallicity simultaneously, gas
accretion of Milky Way (MW) satellite galaxies need to be largely reduced way
before they fall into $R_{Vir}$ of the MW halo, possibly by heating up the
intergalactic medium in the MW halo vicinity to $10^5$K. Alternatively,
\citet{slater13} used simulations to show that environmental effects are
prominent out to 2--3 $R_{Vir}$: satellites with very distant apocenters can be
quenched by tidal stripping and ram pressure stripping following a close
passage to the host galaxy. 

We also extend the local results of G12 to higher \mstar\ by repeating our
measurements on the SDSS sample of \citet[][]{aldo15}. The results (red squares
in Panel (b) of Figure \ref{fig:spatial}), together with G12, suggest that the
fraction of QGs within 2 $R_{Vir}$ has almost no redshift dependence. This
constant fraction suggests that, at all redshifts, environment (especially
within 2 $R_{Vir}$) dominates the quenching of low-mass galaxies.
%{\bf This fraction of QGs between 2 and 4 $R_{Vir}$ has almost no redshift
%evolution (Panel (b) of Figure \ref{fig:spatial}), which suggests that, at all
%redshifts, (1) environment (especially within 2 $R_{Vir}$) dominates the
%quenching of low-mass galaxies and (2) the above three possibilities work on a
%small fraction of galaxies.}

We also study the median distance of galaxies to their massive neighbors scaled
%by $R_{Vir}$ of the massive halos (Panel (c) of Figure \ref{fig:spatial}).
by $R_{Vir}$ (Panel (c) of Figure \ref{fig:spatial}). SFGs have a constant
median distance of $\sim$1.3 $R_{Vir}$ over a wide \mstar\ range. QGs are
closer to massive neighbors, but their median distance depends on \mstar: it
decreases from 1 $R_{Vir}$ at $10^8$\msun\ to 0.5 $R_{Vir}$ at $10^{9.5}$\msun,
then increases to $>1 R_{Vir}$ at $10^{10.5}$\msun. Also, we find no
significant difference between different redshifts.
%two redshift bins in the CANDELS sample as well as no difference between the
%CANDELS and the SDSS samples.

\subsection{Quenching Timescale}
\label{discussion:timescale}

%For quenched galaxies, \dqprojrvir\ allows us to roughly estimate the 
Quenching timescale (\tq) is important to constrain quenching mechanisms. In
the local universe, at \mstar$>10^{10}$\msun, quenching likely occurs through
starvation, whose timescale (4--6 Gyr) is comparable to gas depletion
timescales \citep{fillingham15,peng15}. At \mstar$<10^{8}$\msun, ram pressure
stripping is likely the dominant mechanism
\citep{slater14,fillingham15,weisz15}. Its timescale (2 Gyr) is much shorter
and comparable to the dynamical timescale of the host dark matter halo. 
%There could be a characteristic \mstar, around which 
The dominant quenching mechanism may change around a characteristic \mstar.

To infer \tq, we assume all galaxies start quenching at 4 $R_{Vir}$. We choose
4 $R_{Vir}$ because G12 shows that beyond 4 $R_{Vir}$ the fraction of QGs is
almost zero, while the fraction of SFGs is still high. Theoretically,
\citet{cen14} also predicted the onset of quenching at a similar large halo
distance. Galaxies fall into massive halos while their star formation rates are
being reduced. They become fully quenched when they arrive at the observed
location. Therefore, \tq\ is the time they spent on traveling from 4 $R_{Vir}$
to the observed location with an infall velocity (using circular velocity $V(R)
= \sqrt{ G M(<R) \over R}$ at 2 $R_{Vir}$ as an approximation).

Our method of measuring \tq\ is different from most studies in the literature,
e.g., \citet{wetzel13, wheeler14,fillingham15,balogh16,fossati17}.  They used
numerical simulations or semi-analytic models to match the basic demographics
(e.g., quenched fraction) of QGs. Our method is purely empirical, but relies
on the assumptions of the starting and end points (i.e., 4 $R_{Vir}$ and the
observed location, respectively) of quenching. Our \tq\ definition, however,
characterizes the same physical quantity as other methods, i.e., the timescale
upon which satellites must quench following infalling into the vicinity of
their massive hosts.

The inferred \tq\ is shown in Panel (d) of Figure \ref{fig:spatial}. Overall,
the \tq\ dependence on \mstar\ is, if any, very weak between
$10^8$ and $10^{10}$\msun. Lower-redshift galaxies have longer \tq, because
dynamical timescale decreases with redshift: lower-redshift galaxies need more
time to travel the same \dqprojrvir.

%Because most of other papers study \tq\ as a function of \mstar\ of $z=0$
%galaxies, we take into account of the mass loss of galaxies since the time of
%being quenched to $z=0$. Because for most of \mstar\ bins, the environmental
%quenching can be observed at $z\sim1$, for simplicity, we calculate the mass
%loss of a single stellar population model from $z=1$ to $z=0$. At $z=0$, the
%model only keeps 52\% of its \mstar\ at $z=1$. We therefore shift our
%measurements by -0.28 dex in \mstar\ to compare with others' results.

%Our measurements, even though should be treated as upper limits, Figure
%\ref{fig:timescale} shows that 
Our measurements show excellent agreement with those of \citet{fossati17} and
\citet{balogh16} at \mstar$> 10^{9.5}$\msun\ (Figure \ref{fig:timescale}).
\citet{fossati17} used 3D-HST data
\citep{skelton143dhstcat,momcheva163dhstline} to study the environments of
galaxies with \mstar$\gtrsim 10^{9.5}$\msun\ in CANDELS fields. Agreement with
these detailed studies provides an assurance to our method: although built upon
simplified assumptions, it is able to catch the basic physical principles of
environmental quenching. Moreover, the good agreement also implies that
the projection effect discussed in Section \ref{discussion:spatial} does not
significantly bias our measurement.
%Our results, together with other measurements in the literature, show that
%\tq\ gradually increases from 4 Gyr at \mstar$\sim 10^{8}$\msun\ to 6 Gyr at
%\mstar$\sim 10^{10}$\msun. \tq\ then quickly dropped to 2 Gyr at \mstar$\sim
%10^{11}$\msun. 

%At \mstar$\sim 10^{9.0}$\msun, our measurements are significantly shorter than
%that of \citet{wheeler14} and \citet{fillingham15, fillingham16}. Our shorter
%\tq\ is supported by the fact that the quenching--environment connection (which
%indicates that environmental quenching has already finished for the majority of
%quenched galaxies) is observed at $z\sim1$. Therefore, \tq\ cannot be longer
%than the cosmic time between $z=5$ and $z\sim1$. 

%Overall, together with other measurements in the literature, our results (the
%lower panel of Figure \ref{fig:timescale}) suggest that \tq\ increases
%gradually from 1--2 Gyr at satellite \mstar$\sim 10^{6.5}$\msun\ to 5--6 Gyr
%at \mstar$\sim 10^{9.5}$\msun\ and then drop quickly to 1--2 Gyr at
%\mstar$\sim 10^{11}$\msun.  The trend of \tq\ with satellite \mstar\ is
%critical to understand the quenching mechanisms. 
Our results, together with the measurements of \citet{fossati17} and
\citet{balogh16}, imply a smooth \tq\ transition -- and hence a quenching
mechanism transition -- around \mstar$\sim 10^{9.5}$\msun, which is broadly
consistent with other studies \citep[e.g.,][]{cybulski14,joshualee15}. 
%indicates a change of quenching mechanisms and is broadly consistent with
%other studies \citep[e.g.,][]{cybulski14,joshualee15}. 

At \mstar$\gtrsim 10^{10}$\msun, starvation is likely to be responsible for
environmental quenching \citep{fillingham15}. Alternatively, however, these
galaxies could actually be centrals or recently quenched before becoming
satellites. For them, internal mechanisms (e.g., AGN and star formation
feedback) are likely dominating the quenching, as demonstrated by the
correlation between star formation and internal structures (e.g., central mass
density within 1 kpc discussed in \citet{fang13,barro17,woo17}).

The quenching mechanisms at \mstar$<10^{9.5}$\msun\ are still uncertain. Our
results suggest that \tq\ mildly increases with \mstar\ at $0.5\leq z<0.8$.
Other studies of the local universe
\citep[e.g.,][]{slater14,fillingham15,wetzel15a} suggest a much stronger
\mstar\ dependence of \tq. For example, \citet{fillingham16} argued that \tq\
drops quickly to $\sim$2 Gyr for galaxies with \mstar$\lesssim 10^{8}$\msun\ at
$z\sim0$ because of ram pressure stripping. 
%(\tq$\sim$2 Gyr) may be responsible for quenching low-mass galaxies with
%\mstar$\lesssim 10^{8}$\msun.  Their transition \mstar\ ($\sim 10^{8}$\msun),
%however, is 1.5 orders of magnitude lower than ours.

%Our transition \mstar\ at \mstar$\sim 10^{9.5}$\msun\ is broadly consistent
%with other studies \citep[e.g.,][]{cybulski14,joshualee15}.

At \mstar$> 10^{9.5}$\msun, the redshift dependence of \tq\ can be explained by
the change of the dynamical timescale. We scale up \tq\ at $0.8\leq z<1.2$ by a
factor of $(1+z)^{1.5}$ to account for the redshift dependence of dynamical
timescale (see \citet{tinker10}). This scaled \tq\ (red dashed lines in the
first two panels of Figure \ref{fig:timescale}) matches the actual \tq\
measurements very well at \mstar$\gtrsim 10^{9.5}$\msun. However, it deviates
from the \tq\ measurements at \mstar$<10^{9.0}$\msun. 
%At $z\sim0.65$ (1.7 Gyr after $z\sim1.0$), the deviation is insignificant. But 
At $z\sim0$ (7 Gyr after $z\sim1.0$), the scaled \tq\ is significantly larger
than the \tq\ measured by \citet{fillingham15}. \cite{balogh16} also found
similar results: at $z\sim1$, their \tq\ of galaxies with
\mstar$<10^{10}$\msun\ is longer than the $z\sim0$ \tq\ scaled down by
$(1+z)^{1.5}$. Future work is needed to more quantitatively determine the
redshift dependence of the \tq\ of low-mass galaxies.

\section{Conclusions}
\label{conclusion}

CANDELS allows us to investigate evidence of environmental quenching of dwarf
galaxies beyond the local universe. At $0.5<z\lesssim1.0$, we find that for
$10^{8}$\msun$<$\mstar$<10^{10}$\msun, QGs are significantly closer to their
nearest massive companions than SFGs are, demonstrating that environment plays
a dominant role in quenching low-mass galaxies.
%{\it whether or not} quenched dwarf galaxies live close to massive central
%galaxies at higher redshifts. This quenched dwarf--massive central connection,
%which has been seen in the local universe, is detected in our sample up to
%$z\sim1$, where quenched galaxies are significantly closer to their nearest
%massive companions than star-forming galaxies are at
%$10^{8}$\msun$<$\mstar$<10^{10}$\msun.  , {\bf can be detected up to $z\sim1$
%for $10^{8}$\msun$<$\mstar$<10^{10}$\msun.  beyond the local universe.  the
%connection between quenched dwarf galaxies and massive central galaxies seen
%in the local universe exists at higher redshifts. The evidence of
%environmental quenching, i.e.  We find that quenched galaxies are
%significantly closer to their nearest massive companions than star-forming
%galaxies are, {\bf can be detected up to $z\sim1$ for
%$10^{8}$\msun$<$\mstar$<10^{10}$\msun. 
We also find that about 10\% of the QGs in our sample are located between two
and four $R_{Vir}$ of the massive halos. 
%implying that quenching starts from large halo radii. 
The median projected distance from the QGs to their massive neighbors
($d_{proj}^Q/R_{Vir}$) decreases with satellite \mstar\ at \mstar$\lesssim
10^{9.5}$\msun, but increases with satellite \mstar\ at \mstar$\gtrsim
10^{9.5}$\msun. This trend suggests a smooth, if any, transition of \tq\ around
\mstar$\sim 10^{9.5}$\msun\ at $0.5<z<1.0$.

%depends on satellite \mstar, but not
%on redshift. The \tq\ inferred from $d_{proj}^Q$ suggest a smooth, if any,
%transition of quenching mechanisms around \mstar$\sim 10^{9.5}$\msun\ at
%$0.5<z<1.0$.
%
%}

%\ and (2) no change of quenching mechanism between $z \sim 1$ and $z=0$.
%but cannot be detected at $z>1.2$. The
%highest $z$ to which such an evidence can be found slightly depends on
%\mstar\ of the low-mass galaxies. Our method provides a uniform way to
%constrain \tq\ over two orders of magnitude in satellite \mstar. We find that
%the upper limit of \tq\ gradually increases from 4 Gyr at \mstar$\sim
%10^{8.5}$\msun\ to 6 Gyr at \mstar$\sim 10^{10.5}$\msun. Our results are
%broadly consistent with other studies of \tq\ at \mstar$\sim 10^{8}$\msun\ and
%at \mstar$\sim 10^{10}$\msun. Together, they imply a transition of \tq\ (and
%therefore quenching mechanisms) at \mstar$\sim 10^{10}$\msun.
%At $z\lesssim1$ and $10^{8}$\msun$<$\mstar$<10^{10}$\msun, the $d_{proj}$
%distributions of quenched galaxies are significantly skewed toward lower
%values than those of star-forming galaxies. Such a difference between the two
%populations, however, disappears at $z>1.2$. This transition around $z=1$
%places a constraint on the environmental quenching timescale (\tq). We find
%that \tq\ gradually increases from 4 Gyr at \mstar$\sim 10^{8.5}$\msun\ to 6
%Gyr at \mstar$\sim 10^{10.5}$\msun. Our method provides a uniform way to
%constrain \tq\ over two orders of magnitude in satellite \mstar.

\

Y.G., D.C.K., and S.M.F. acknowledge support from NSF grant AST-0808133.
Support for Program HST-AR-13891 and HST-GO-12060 were provided by NASA through
a grant from the Space Telescope Science Institute, operated by the Association
of Universities for Research in Astronomy, Incorporated, under NASA contract
NAS5-26555. Z.C. acknowledges support from NSFC grants 11403016 \& 11433003.
% MR acknowledges support from an appointment to the NASA Postdoctoral Program
% at Goddard Space Flight Center.

% {\it Facilities}: {\it HST} (WFC3), Keck (DEIMOS)

%\begin{thebibliography}{1}
%\expandafter\ifx\csname natexlab\endcsname\relax\def\natexlab#1{#1}\fi
%
%\bibitem[{{Oke}(1974)}]{oke74}
%{Oke}, J.~B. 1974, \apjs, 27, 21
%
%\end{thebibliography}

%\appendix

%\twocolumngrid

%\bibliographystyle{apj}
%\bibliography{references}
%\input{clumpprp_v0a.bbl}
%\newpage
%\small
%\input{./table_glx.tex}
%\input{./table_clump_mcerr.tex}
%\input{./table_comp.tex}

\end{document}